\documentclass[12pt]{article}
\usepackage{graphicx,amsmath,amssymb,color,bm}

\begin{document}

\centerline {\Large\textbf {Combined Effect of Stacking and Magnetic Field  }}
\centerline {\Large\textbf {on Plasmon Excitations in Bilayer Graphene}}


\centerline{Jhao-Ying Wu$^{1,\star}$, Godfrey Gumbs$^{2,3,\dag}$, and Ming-Fa Lin$^{1,\dag\dag}$ }
\centerline{$^{1}$Department of Physics, National Cheng Kung University,
Tainan, Taiwan 701}
\centerline{$^{2}$Department of Physics and Astronomy, Hunter College at the City University of New York,\\ \small }
\centerline{695 Park Avenue, New York, New York 10065, USA}
\centerline{$^{3}$Donostia International Physics Center (DIPC),
P de Manuel Lardizabal, 4, 20018}
\centerline{San Sebastian, Basque Country, Spain}

\vskip0.6 truecm

\noindent

The electronic excitations of bilayer graphene (BLG) under a magnetic field are investigated with the use of  the Peierls tight-binding model in conjunction with random-phase approximation (RPA). The interlayer atomic interactions, interlayer Coulomb interactions, and magnetic field effects are simultaneously included in the dielectric-function matrix. That enables us to derive the magneto-Coulomb-excitation spectrum of different stacking structures. The two typical arrangements of BLGs, AB and AA, are considered in this article. AB-BLG exhibits many discrete energy-loss peaks, which correspond to  the quantization of electron energies. On the other hand, the AA-BLG spectra possess a unique and pronounced peak at low frequency. This peak represents the collective excitation of the entire low-frequency Landau states. The dependence of the energy-loss peaks on the momentum transfer and the magnetic field strength is presented. Accordingly, two kinds of plasmon modes produced by the layer stacking are clearly distinguished.

\vskip0.6 truecm

$\mathit{PACS}$: 73.21.Ac, 73.22.Lp, 73.22.Pr

\newpage

\bigskip

\centerline {\textbf {I. INTRODUCTION}}%

\bigskip

\bigskip

Monolayer graphene (MLG) is well known for its unusual linear band structure at low energy near the K and K$^\prime$ points \cite{Wallace:1947}. This characteristic behavior gives rise to several unusual physical properties. These include the Klein paradox for tunneling through a potential barrier\cite{Katsnelson:2006,Bai:2007,Young:2009}, the anomalous quantum Hall effect\cite{Zhang:2005,McCann:2006}, and peculiar optical properties \cite{Abergel:2007,Nair:2008,Bonaccorso:2010,HoYH:2010,Zheng:2012}. However, achieving practical applications such as a field-effect transistor \cite{Zhang:2009,Schwierz1:2010} as well as graphene interconnects \cite{Sordan:2009,Yang:2010,Chen:2010} relies on tunable electronic properties. One practical way to achieve the tunability is by stacking graphene layers. The idea is based on the strong dependence of the energy bands on the stacking order, the number of layers, or the use of external fields.

Bilayer graphene (BLG) is the simplest and the most widely studied few-layer graphene system. It can be fabricated by mechanical exfoliation \cite{Zhang:2005,Novoselov:2004,Geim:2007,Miao:2007} or epitaxial growth \cite{Lee:2008}. These two methods are considered feasible for producing AB- and AA-stacking structures, respectively. The A and B atoms (the two sublattices of the hexagonal structure of graphene) in the AB configuration have different chemical environments. This causes the breakdown of the hexagonal symmetry. The energy bands thus exhibit parabolic-like energy dispersion \cite{Lu:2006,Partoens:2006}. On the other hand, two graphene layers in an AA-stacked configuration are stacked in a highly symmetric way normal to the planes. This causes the low-energy bands to look like a collection of those of MLG, i.e., two pairs of linear bands. The dissimilar band structures of AA- and AB-stacked graphenes are the main  reason for their distinct physical phenomena, e.g., their optical properties \cite{Abergel:2007,HoYH:2010,Chang:2008,HOYH2:2010} and Coulomb excitations \cite{Ho:20061,Chuang:2013}. The Coulomb excitations, associated with the screened Coulomb potential, can be used to investigate the collective-excitation properties. It was found that hardly any low-frequency plasmons exist in AB-BLG due to the insufficient number of free carriers. However, a strong acoustic plasmon is predicted for AA-BLG \cite{Ho:20061}.

The plasmon behavior in BLG may be further tuned by external fields. A perpendicular electric field can produce a band gap in AB- \cite{Zhang:2009,Chuang:2013,Ohta:2006,Zhou:2007} and increase the number of free carriers in AA-BLGs \cite{Lin:2012}. Consequently, the E-field brings about well-defined optical plasmons in AB- \cite{Chuang:2013} and enhances the acoustic plasmon in AA-BLGs \cite{Lin:2012}. On the other hand, a magnetic field can quantize the electron energies and induce two groups of Landau levels (LLs) in BLGs \cite{HOYH2:2010,Lai:2008} as a result of the interlayer atomic interactions. A number of studies investigate magnetoplasmons in monolayer graphene \cite{Berman:2008,Roldan:2009,Wu:2011,Roldan:2011}. The plasmon dispersions reflect the strong competition between the Lorentz force and the Coulomb interaction \cite{Berman:2008,Wu:2011}. To our knowledge, a comprehensive study on the behavior of magnetoplasmons under the stacking effects and interlayer Coulomb interactions has not yet been carried out to date.

This study is devoted to the calculation of magneto-Coulomb excitation properties of AB- and AA-BLGs. With the Peierls tight-binding model, the complete $\pi$-band LL structure and wavefunctions are obtained. The dielectric function calculated within the random-phase approximation  (RPA) is cast into   matrix form to include both, the interlayer atomic interactions and interlayer Coulomb interactions. Single- and many-particle excitation spectra are presented. Two kinds of bare polarization functions exist, namely an intralayer and an interlayer function. Whether these functions are in phase or not depends on their frequency and the graphene stacking type. The energy-loss function is employed to investigate the collective excitations. Our calculations have shown that AB-BLG exhibits many discrete inter-LL plasmons that reflects the zero-dimensional features of energy bands. On the other hand, AA-BLG possesses a unique 2D-like plasmon, which arises from the highly symmetric stacking order responsible for the dense LLs around the Fermi level. The significant difference in plasmon behaviors between AB- and AA-BLGs demonstrates the important role played by the interlayer atomic interactions. The presented results may be further validated by inelastic light-scattering \cite{Gumbsbook}, which has been successfully applied to the two-dimensional electron gas system in magnetic fields \cite{Eriksson:1999,Richards:2010}.

\bigskip
\bigskip
\centerline {\textbf {II. METHODS}}%
\bigskip
\bigskip

AB-stacked BLG consists of two coupled monolayers, indexed 1 and 2,  with an interlayer
separation $d$=3.35 ${\AA}$. Each graphene sheet is composed of sublattices A and B and the C-C bond
length is $b$=1.42 ${\AA }$. If the atoms are labeled $A^{1}$ and
$B^{1}$ in sheet 1 and $A^{2}$ and $B^{2}$ in sheet 2, then the
$A^{2}$ atoms are stacked directly over the $A^{1}$ atoms, and the $B^{2}$
atoms are projected onto the hexagonal centers of the adjacent
layer. Four kinds of atom-atom interactions have to be taken into account for calculations \cite{Nakao:1976}. $\alpha_{0}$=3.12 eV, $\alpha_{1}$=0.38 eV,
$\alpha_{3}$=0.28 eV, and $\alpha_{4}$=0.12 eV are the atomic hopping
integrals, respectively, between two nearest-neighbor atoms, $A^{1}$
and $A^{2}$ atoms, $B^{1}$ and $B^{2}$ atoms, and $A^{1}$ and
$B^{2}$ atoms ($A^{2}$ and $B^{1}$ atoms). The difference in site
energy between the nearest-neighbor atoms is characterized by
$\alpha_{6}$=-0.0366 eV.

The geometric structure of AA-stacked bilayer graphene is that all the carbon atoms on the two sheets are stacked directly with an interlayer
distance of \emph{c}=3.35 {\AA}. The C-C interactions are $\alpha_{0}$=2.569 eV, $\alpha_{1}$=0.361 eV, and $\alpha_{3}$=-0.032 eV \cite{Khodkov:2012}, which respectively represent the atomic hopping integral between the nearest-neighbor atoms on the same layer, the atoms \emph{A} or \emph{B} from the nearest-neighbor
layers, and the atoms \emph{A} and \emph{B} from two nearest-neighbor layers. The first Brillouin zone of BLG with symmetric points $\Gamma$(0,0), M(2$\pi$/(3\emph{b}),0) and
K(2$\pi$/(3\emph{b}),2$\pi$/(3$\sqrt{3}$\emph{b})) is the same as
that of a graphene sheet.

In the tight-binding model, the crystal Hamiltonian can be represented by the Bloch functions of the four periodic
carbon atoms $|A^{1}\rangle$, $|B^{1}\rangle$, $|A^{2}\rangle$ and
$|B^{2}\rangle$. Only the $2p_{z}$ orbitals are taken into
account when investigating the low-energy electronic structures. The BLG is assumed to be under the influence of a uniform magnetic
field $\mathbf{B}=B\hat{z}$ perpendicular to the graphene plane. The magnetic flux, a product of the
field strength and the hexagon area, is $\Phi =
[3\sqrt{3}b^{2}B/2]/\phi_{0}$ in units of the flux quantum
$\phi_{0}$ = $hc/e$ = 4.1356¡Ñ1015 T/$m^{2}$. The vector potential, which is chosen as \textbf{A} =
$(Bx)\hat{y}$, leads to a new periodicity along the
armchair direction \cite{Lai:2008}. The unit cell is thus enlarged and its
dimension is determined by $R_{B}= 1/\Phi$. The enlarged unit cell contains 8$R_{B}$ carbon atoms (for a rectangular unit cell) and the Hamiltonian matrix is a 8$R_{B}$$\times$8$R_{B}$ Hermitian matrix. The first Brillouin zone is extended to
the range of $|k_{x}|<\pi/3bR_{B}$ and $|k_{y}|<\pi/\sqrt{3}b$. That is, the number of charge carrier per area is self conserved.

For $B$=40 T, the Hamiltonian matrix is a
$8000\times8000$ matrix. When the field strength descends to $B$=5 T, the size of matrix becomes $64000\times64000$. To solve such a huge matrix, the base functions are rearranged to let the nonzero matrix elements locate around the diagonal line to form a band-like matrix \cite{Lai:2008}. Therefore, all eigenvalues in the energy range ($\pm5$ eV) related to $\pi$-electron can be evaluated efficiently. The distribution of wave function in real space can be clearly demonstrated, so the quantum number of each LL is well defined by the node number of its wave function. Also, the degeneracy of LL is calculated.

When the system is perturbed by a time-dependent electrostatic potential, the electron-electron interactions would induce a charge density fluctuation which acts to screen the external perturbation. This dielectric screening determines the normal modes of the charge density oscillations. In bilayer graphene, charges in the two layers would participate in the screening. Within the random-phase approximation (RPA), the $2\times 2$ dielectric-function matrix can be written as: \cite{Ho:20062}

\begin{equation}
\left[ \begin{array}{cccc}
            \epsilon_{11}(\textbf{q},\omega) & \epsilon_{12}(\textbf{q},\omega) \\
            \epsilon_{21}(\textbf{q},\omega) & \epsilon_{22}(\textbf{q},\omega)
          \end{array}
       \right]
=\left[
          \begin{array}{cccc}
            \epsilon_{0}-\sum_{l}V_{1l}(q)P_{l1}^{(1)}(\textbf{q},\omega) & -\sum_{l}V_{1l}(q)P_{l2}^{(1)}(\textbf{q},\omega) \\
            -\sum_{l}V_{2l}(q)P_{l1}^{(1)}(\textbf{q},\omega) & \epsilon_{0}
-\sum_{l}V_{2l}(q)P_{l2}^{(1)}(\textbf{q},\omega) \\
          \end{array}
       \right],
\end{equation}
with $\epsilon_{0}$=2.4 being the background dielectric constant. The bare
Coulomb potentials for interlayer and intralayer are
$V_{ll'}(q)=v_{q}e^{-q|l-l'|c}$, where $v_{q}=2\pi$$e^2/q$. In the presence of a magnetic filed, the system becomes
fully quantized. The interlayer and intralayer response function may be expressed as:

\begin{align}
P_{ll'}^{(1)}(\textbf{q},\omega)&=2\sum_{n,n';k}(\sum_{i}u_{nli}
(\textbf{k})u_{n'li}^{*}(\textbf{k}+\textbf{q}))\notag
\\&\times(\sum_{i'}u_{nl'i'}^{*}(\textbf{k})u_{n'l'i'}(\textbf{k}+\textbf{q}))\notag
\\&\times\frac{f(E_{n}(\textbf{k}))-f(E_{n'}(\textbf{k}+\textbf{q}))}
{E_{n}(\textbf{k})-E_{n'}(\textbf{k}+\textbf{q})+\hbar\omega+i\delta}.
\end{align}
\emph{f}($E_{n}$(\textbf{k}))=1/[1+exp($E_{n}$(\textbf{k})-$\mu$(T))/($k_{B}$T)]
is the Fermi-Dirac distribution, where $\mu$(T) is the
temperature-dependent chemical potential and $k_{B}$ is Boltzmann's
constant. The crystal wavefunction $\Psi_{n}(k)=\sum_{li}u_{nli}(k)U_{li}(k)$, where $l$=1,2 (the layer number) and $i=a_{1}, b_{1}, a_{2}, b_{2}....a_{R_{B}}, b_{R_{B}}$ (the $2R_{B}$ sublattices). $u_{nli}$ are the respective coefficients of the $4R_{B}$ tight-binding functions ($U_{li}$). $E_{n}$ is the eigenvalue corresponding to the eigenfunction $\Psi_{n}$. Only $q_{y}$ and $k_{y}$ components are considered here. The
calculation along the other direction of $q_{x}$ and $k_{x}$ leads to the
same results. The interlayer polarizations ($l\neq l'$) vanish when the
interlayer atomic interactions are neglected.

\par
The energy-loss function, or the inelastic scattering probability, is evaluated from Born approximation and Fermi's golden rule. The details of the evaluation can be found in Ref. \cite{Ho:20062}. The final form is written as:

\begin{align}
\textrm{Im}[-1/\varepsilon]\equiv\sum_{l=1,2}\textrm{Im}[-V^{eff}_{ll}(\mathbf{q},\omega)]
/(\sum_{ll'}V_{ll'}(q)/\emph{2}) \ ,
\end{align}
where the effective Coulomb potential is defined as $[V^{eff}]=[\epsilon]^{-1}[V]$.
The denominator is the average of bare Coulomb potentials on all layers. The screened response function is used to understand the collective excitations of the low-energy $\pi$ electrons.

\bigskip
\bigskip
\centerline {\textbf {III. RESULTS AND DISCUSSION}}%
\bigskip
\bigskip

In Fig. 1, we present the low-energy bands of BLGs and their corresponding LL spectra under a magnetic field of \emph{B} = 40 T. At \emph{B} = 0, AB-stacked BLG owns two pairs of parabolic bands, as denoted by the different colors in the left panel of Fig. 1(a). They are primarily dominated by A and B atoms, respectively, as a result of the different chemical environments of the two sublattices. A small band overlap exists around the Fermi level (inset of Fig. 1(a)). The overlap creates a number of free carriers, and thus AB-BLG is considered a semimetal. Under an external magnetic field, the two pairs of parabolic bands evolve into the two groups of LLs with a respective series of quantum numbers ($n^{c,v}$ = 0,1,2,3...), as indicated in the right panel of Fig. 1(a) by different colors . The first group begins at the Fermi level (black lines), and the second group (red lines) occurs at energy higher than the strongest interlayer-interaction term $\alpha_{1}$ ($\sim$0.38 eV, see the {\bf Methods} section). The $n^{c}$ = 0 and $n^{v}$ = 1 LLs in the first group determine the band gap. The opening of the gap is associated with the band overlap that exists at $B$ = 0. The almost equal LL spacings at low energies ($E<\alpha_{1}$) are attributed to the parabolic energy dispersion. However, at $E>\alpha_{1}$, the coexistence of the two LL groups causes the highly unequal energy spacings. The LL spectrum is slightly asymmetric about the Fermi level due to interlayer interaction $\alpha_{3}$ (the interaction between B atoms from different layers). At $B$ = 0, $\alpha_{3}$ leads to the nonidentical band slopes of the valence and conduction bands.

For AA-BLG (Fig. 1(b)), two pairs of linear bands exist at low energy. In the presence of a magnetic field, the two pairs evolve into two groups of LLs, both of which obey the energy relation $E_{n}\propto\sqrt{nB}$  but with different proportionality constants ($n$ is the quantum number in each group). The two $n=0$ LLs indicated in the figure correspond to the intersections of each pair  of linear bands. The energy gap is constructed by two LLs from different groups (at $B$ = 40 T these are $n^{c}=4$ and $n^{v}=4$ LLs). When the field strength is varied, the energy spacing, the amount of LLs, and the quantum numbers of the lowest unoccupied and the highest occupied LLs change accordingly, leading to the observed oscillatory behavior of the band gap size \cite{Tsaia:2012}.

The intralayer polarization functions of AB- and AA-BLGs at different momentum transfers ($q$ = 10 and $q$ = 30 in units of $10^{5}$ $cm^{-1}$) under $B$ = 40 T are shown in Fig. 2. Each symmetric peak (pair of asymmetric peaks) in the imaginary (real) part represents a principal inter-LL transition channel that obeys the momentum and energy conservation laws. Their intensity is determined by the wavefunction overlap between the initial and final states. In this figure, the most prominent peaks are marked by the pair ($n^{v}$,$n^{c}$) which denotes the quantum numbers of the initial and final states, respectively. The peak distributions show a strong dependence on the momentum transfer and the graphene stacking order. In AB-BLG (Figs. 2(a) and 2(b)), the first-group LL transitions dominate the low-frequency spectrum. Peaks with small transition orders (as determined by $\triangle n=|n^{v}-n^{c}|$) are especially strong at small $q$. Some twin-peak structures are observed, whose existence is the result of three different factors. The first is the asymmetric LL spectrum about the Fermi level. This leads to two equally high peaks, e.g., (2,3) and (3,2). The second is that two transition channels happen to have similar energies, which induces two adjacent peaks with distinct intensities and dependence on the momentum, e.g. peaks (2,2) and (1,4) in Fig. 2(b). The third factor is that one peak is associated with $n^{c}$ = 0 LL and the other with $n^{v}$ = 1, e.g., (1,2) and (0,2). These two peaks have similar energies but different intensities. The single-particle excitation (SPE) channels with a larger transition order need a larger momentum to be triggered; for instance, peak (1,4) is absent at $q$ = 10 but occurs at $q$ = 30. On the contrary, those with a smaller transition order are suppressed at larger $q$ (e.g. (2,2) and (3,3)). This reflects the features of the Hermite polynomial functions. At frequencies higher than $2\alpha_{1}$, the second-group LL transitions are observed, e.g., peak (0,0). Transitions between the first and the second LL groups are weak since the two groups of LLs are dominated by different sublattices and the wavefunction overlap between them are quite small.

In AA-BLG (Figs. 2(c) and 2(d)), the low-frequency peaks ($\omega<2\alpha_{1}\sim0.72$ eV) are dominated by the intraband ($n^{v}\rightarrow n^{v}$ and $n^{c}\rightarrow n^{c}$) and intragroup LL transitions. At $q=10$ (Fig. 2(c)), there is a wide flat region (no peaks) between $0.1<\omega<0.5$ eV. This manifests the small transition probability for larger transition orders of intragroup LL transitions. This frequency region is later shown to contain a strong plasmon peak involving the entire low-energy Landau states. At a frequency of around $2\alpha_{1}$ there is a sharp peak co-contributed by channels (0,0), (1,1), (2,2), and (3,3), which represents intergroup LL transitions with the same quantum numbers. This is in large contrast to AB-BLG, in which the intergroup LL transitions are faint. Above a frequency of $2\alpha_{1}$, the peaks of intergroup and intragroup transitions can coexist at some momentum transfers.

The Peierls tight-binding Hamiltonian matrix is based on the enlarged unit cell that fits the magnetic-field-induced periodic boundary condition. The distribution of real-space Landau wavefunctions shows that their localization positions vary with the wavevector $k$, and the localization width becomes wider as the quantum number increases (the details can be seen in Ref. 30). At a smaller momentum transfer $q$, even the initial and final states with the narrower distribution widths (corresponding to the smaller quantum numbers) can have a significant overlap. However, at larger momentums, it needs at least one of the two state wavefunctions with the wider localization width to have the nonzero overlap, because the two states with the larger $k$ difference between them have the farther separated wavefunction localization positions. This generally explains that why peaks with the smaller transition orders (as determined by $\triangle n=|n^{v}-n^{c}|$) are especially strong at a smaller $q$, and the opposite is true for a larger $q$.

There is some difference between the intralayer ($P_{11}^{(1)}$) and interlayer ($P_{12}^{(1)}$) polarization functions. A comparison between them is presented in Fig. 3. For AB-BLG, imaginary parts Im[$P_{11}^{(1)}$] and Im[$P_{12}^{(1)}$] exhibit different peak heights at lower frequencies (Fig. 3(a)). When $\omega$ is increased, the intensity difference is reduced, while Im[$P_{11}^{(1)}$] and Im[$P_{12}^{(1)}$] begin to have opposite signs from each other, i.e., they become out-of-phase. The different intensity at lower frequency is caused by the unequal contributions of A and B atoms to the two layers, and the different sign at higher frequency is a direct consequence of interband transitions when charge distributions are equal (or almost equal) on the two layers. The absence of peak (0,1) in Im[$P_{12}^{(1)}$] is attributed to the fact that the wavefunction of $n=0$ LL is confined to only one graphene layer. For AA-BLG, Im[$P_{11}^{(1)}$] and Im[$P_{12}^{(1)}$] have almost the same weights in the whole frequency region, as shown in Fig. 3(b). The two polarization functions are in phase at $\omega<0.3$ eV where the intraband LL transitions dominate the excitation spectrum. At $\omega>0.3$ eV, the intergroup and interband transitions are responsible for the out-of-phase behavior. A comparison between real parts Re[$P_{11}^{(1)}$] and Re[$P_{12}^{(1)}$] is presented in Figs. 3(c) and 3(d) for AB- and AA-BLGs, respectively. In both stacking types, Re[$P_{11}^{(1)}$] and Re[$P_{12}^{(1)}$] in the in-phase region show a relative shift in the vertical direction. This vertical shift is produced by the out-of-phase behavior at higher frequency ($\omega>0.3$ eV). It causes the two functions to have their zero points at different positions. The above discussions suggest that the intralayer and interlayer charge scatterings are distinguishable and that screening effects from higher LL transitions are significant. In other words, appropriate interlayer charge distributions and a wide energy range of LLs are necessary when calculating the screened Coulomb interactions in a layered graphene system.

The energy-loss function Im$[-1/\varepsilon]$ is employed to investigate the collective excitations. In AB-BLG, it exhibits many discrete peaks at lower frequencies, shown in Figs. 4(a) and 4(b) for different momentum transfers. These discrete peaks reflect the zero-dimensional features of LLs and are considered inter-LL plasmons. Their existence, frequency, and height strongly depend on the momentum transfer. The first peak remains much more stable than the others when $q$ is varied. It is mainly dominated by the band-gap transition. The band-gap transition is a result of the asymmetric stacking order; it is absent in AA-BLG and monolayer graphene.

The energy-loss function of AA-BLG displays distinct features. It possesses a unique and prominent peak at low frequency (Figs. 4(c) and 4(d)). This peak can be regarded as a 2D-like plasmon because it exhibits the involvement of all low-frequency Landau states and a nearly square-root dispersion versus $q$ (shown in Fig. 5(b)). The densely packed LLs around the Fermi level are the main cause of this. Furthermore, several discrete peaks are located at the right side of the 2D-like plasmon. Those peaks correspond to the sparse interband and intergroup LL transitions and are categorized as inter-LL plasmons. The weight of the inter-LL plasmons is increased by raising $q$. Meanwhile, the 2D-like plasmon moves to a higher frequency with its intensity reduced due to the enhanced Landau damping. The coexistence of the two plasmon modes results from the unequally spaced LLs and is a novel feature of AA-BLG. Although the intergroup LL transitions lead to the strong divergent peaks in single-particle polarization functions (Figs. 2(c) and 2(d)), they do not induce pronounced plasmon peaks. The significant Landau damping from the interband and intragroup LL transitions is responsible for this.

The peak frequencies of Im$[-1/\varepsilon]$ as a function of $q$ are presented in Fig. 5. The inter-LL plasmons in AB-BLG show discrete branches (Fig. 5(a)). Each has a minimum excitation energy approaching its dominant SPE energy in both the short and long wavelength limits where the polarization shift may be neglected. The peculiar plasmon dispersion arises from the strong competition between the Lorentz force and Coulomb interaction \cite{Wu:2011}. The slope of each mode is decided by the dominant Landau state wavefunctions. Therefore, two SPE channels with close energies and similar state wavefunctions may co-contribute to one branch --- (1,2) and (0,2) indicated in the figure for example. Such a composite plasmon mode begins and ends at different SPE energies and thus has a longer lifetime than that of other single modes. If two SPE channels have a similar energy but distinct state wavefunctions, they may cause an oscillatory plasmon dispersion, e.g., the branches marked by (2,2) and (1,4). These peculiar features are hard to see in monolayer graphene for low frequencies due to the smaller number of LLs and the symmetric e-h bands \cite{Wu:2011}.

The plasmon dispersion in AA-BLG  differs substantially from that in AB-BLG. The intraband LL transitions ($n^{c}\rightarrow n^{c}$ and $n^{v}\rightarrow n^{v}$) cause a unique plasmon branch at low frequency (Fig. 5(b)), which is a mixture of the cyclotron mode and the usual 2D plasmon with a $\sqrt{q}$ dispersion relation. The existence of a gap at $q=0$ is mainly an effect of the magnetic field. The 2D-like dispersion requires a sufficient density of LLs around the Fermi level. In AA-BLG, this density is naturally induced by the interlayer atomic interactions. When $q>20$ (frequency higher than the strongest coupling term $\alpha_{1}\sim0.361$ eV), the interband ($n^{v}\rightarrow n^{c}$ and $n^{c}\rightarrow n^{v}$) and intergroup LL transitions gradually replace the intraband transitions and dominate the excitation spectrum. Then, discrete branches representing the inter-LL plasmons occur. The transformation from a 2D-like plasmon to inter-LL plasmons with the change of the momentum transfer is attributed to the unusual LL distribution caused by the highly symmetric stacking order.

The plasmon dispersion versus the strength of the magnetic field is plotted in Fig. 6. In AB-BLG (Fig. 6(a)), the intensity and frequency of inter-LL plasmons increase monotonically with the field strength, which corresponds to the enhanced degeneracy and the expanded spacings of LLs. There is a critical value of $B$ above which the band-gap transition occurs. The energy gap is further widened when $B$ is increased, which is reflected in a rise of the peak frequency. It is noted that some plasmon modes break their continuity at certain field strengths. At those points, twin-peak structures are observed in the energy-loss function. This mainly comes from the asymmetry of electron and hole bands. The asymmetry is weakened at stronger fields, and consequently the discontinuities disappear above $B=30$ T. In the case of AA-BLG (Fig. 6(b)), the 2D-like plasmon, which makes the most contributions to the spectrum at low frequency, starts around $\omega=0.4$ eV when $B\rightarrow$ 0. Its frequency oscillates with $B$ because of the rapid change in the LL quantum numbers around the Fermi level. This 2D-like plasmon may evolve into several inter-LL plasmons at the region $B>40$ T, where the number of LLs is significantly reduced.

The Peierls tight-binding model calculations presented in this manuscript are more complete than the continuous band approximation (the effective-mass model) when investigating the Coulomb excitations in layered graphene. First, the whole $\pi$-band Landau level structure is obtained and the number of carriers per area is self-conserved. Therefore, the accuracy of the magnetoplasmons is not hindered by either the field strength or the energy range. However, in the continuous band approximation, it needs a field-strength-dependent cutoff of Landau levels (LLs) to conserve the carrier number and to ensure the validity of the energy bands. This cutoff of LLs is hard to define. Second, the Peierls tight-binding Hamiltonian matrix includes the field effects and all meaningful interlayer atomic interactions simultaneously. Therefore, it offers the more correct LL spectrum and charge distributions on all layers. The latter directly decides the weights of intralayer and interlayer polarization functions, as shown in Fig. 3. In the effective-mass model calculations, some interlayer hopping terms are dropped to simplify the energy-band hybridizations at low energies. Therefore, the intralayer and interlayer polarization functions obey the simple relation $P_{11}^{(1)}=P_{12}^{(1)}$ or $P_{11}^{(1)}=-P_{12}^{(1)}$ in the whole energy region \cite{Borghi:2009,Triola:2012}, where 1 or 2 is the layer index. In other words, the charge oscillations are either totally in phase or out of phase. However, in Fig. 3, we point out that the weights of $P_{11}^{(1)}$ and $P_{12}^{(1)}$ could be equal or unequal depending on the frequency and the layer stacking order, which demonstrates the importance of considering all meaningful interlayer atomic interactions. Third, there are two pairs of energy bands in bilayer graphene, and single-particle excitations (SPEs) are allowed in the same pair or between different pairs. Both transition channels contribute to the screening effects during the Coulomb excitation process and are completely included in the presented calculations. However, in the effective-mass model calculations, the screening effects from the different kinds of transition channels are not treated simultaneously. The calculation becomes more complex when a magnetic field is applied. Both intragroup and intergroup LL transitions induce prominent peaks in the single-particle polarization functions, as shown in Fig. 2. The two kinds of transition channels may coexist in the same frequency region and co-contribute to magnetoplasmons or Landau damping. Moreover, Fig. 3 shows that the out-of-phase charge oscillations at higher frequency may cause Re[$P_{11}^{(1)}$] and Re[$P_{12}^{(1)}$] to have a relative shift in the vertical direction and make them have different zero points. That suggests the importance of the screening effects from the higher-energy state transitions.

\bigskip
\bigskip
\centerline {\textbf {IV. SUMMARY AND CONCLUSIONS}}%
\bigskip
\bigskip

We investigate the Coulomb-excitation properties of BLGs under magnetic fields by using the Peierls tight-binding model and RPA. The interlayer atomic interactions, interlayer Coulomb interactions, and field effects are simultaneously included in the dielectric-function matrix. By utilizing the matrix, the energy-loss function is obtained, which is then used to determine the plasmon excitations. In AB-BLG, the energy-loss spectrum is mainly dominated by discrete inter-LL transitions, while in AA-BLG, a 2D-like plasmon involving the entire low-frequency Landau states is found. The latter is a result of the highly symmetric stacking order and the dense LL distribution around the Fermi level. The inter-LL plasmon and the 2D-like plasmon have a very different dependence on the momentum transfer and the magnetic field strength. Their behaviors are well understood from the fundamental single-particle-excitations and the screening effects in the polarization functions. The results presented in this article suggest that tunable collective-excitation properties in graphene may be achieved by the combined effects of the stacking order and the magnetic field. Doping the system is able to control the number of occupied LLs and may lead to a significant change of low-frequency peak distributions in AB BLG. The 2D-like plasmon in AA BLG would remain stable in the doped conditions, and a few of inter-LL plasmons may be enhanced. The relevant formulas and the analysis can be easily extended to a larger number of layers or bulk graphite under any form of external fields. Therefore, accurate knowledge of layered graphene systems and their possible applications in plasmonics may be acquired on the basis of this study.

\bigskip

\bigskip

\centerline {\textbf {ACKNOWLEDGMENT}}%

\bigskip

\bigskip

\noindent \textit{Acknowledgments.} This work was supported by the NSC of Taiwan, under Grant No. NSC 98-2112-M-006-013-MY4.

\newpage

\par\noindent ~~~~$^\star$e-mail address: yarst5@gmail.com

\par\noindent ~~~~$^\dag$e-mail address: ggumbs@hunter.cuny.edu

\par\noindent ~~~~$^\dag$$^\dag$e-mail address: mflin@mail.ncku.edu.tw


\newpage

\centerline {\Large \textbf {Figure Captions}}

\vskip0.3 truecm

Figure 1. The energy bands of (a) AB- and (b) AA-BLGs at zero field
and in the presence of a magnetic field \emph{B}= 40 T. The different colors denote the different pairs of energy bands or different groups of LLs. A few of LLs are indicated by $n^{c}$ and $n^{v}$, where $n$ is the quantum number and $c$ and $v$ represent the conduction and valence bands, respectively. The inset in (a) shows the band overlap at $B$ = 0 in AB-BLG.

\vskip0.5 truecm

Figure 2. The real (dashed curve) and imaginary (solid curve) parts of the single-particle intralayer polarization functions of AB-BLG are shown in (a) and (b) for $q$ = 10 and $q$ = 30, respectively, where $q$ is in unit of $10^{5}$ $cm^{-1}$. The same plots for AA-BLG are shown in (c) and (d). The most prominent peaks are labeled by pairs of LL quantum numbers $(n^{v},n^{c})$ in the same or different colors, which represent the intragroup and intergroup LL transitions, respectively. The energy-broadening parameter is $\delta$=5 meV.
\vskip0.5 truecm

Figure 3. This figure shows the comparison between the intralayer and interlayer polarization
functions for their imaginary parts in (a) and (b) and real parts in (c) and (d)(for AB ((a) and (c)) and AA-BLGs ((b) and (d))).

\vskip0.5 truecm

Figure 4. The energy loss function at $q$=10 and 30 for AB- ((a) and (b)) and
AA-BLGs ((c) and (d)).

\vskip0.5 truecm

Figure 5. The momentum dependence of the plasmon frequencies in (a) AB- and (b)
AA-BLGs.

\vskip0.5 truecm

Figure 6. The dependence of the plasmon frequencies on the strength of the magnetic field for (a) AB- and (b) AA-BLGs at $q$=20.

\newpage

\begin{figure}
  \begin{center}
  \includegraphics[height=0.5\textheight]{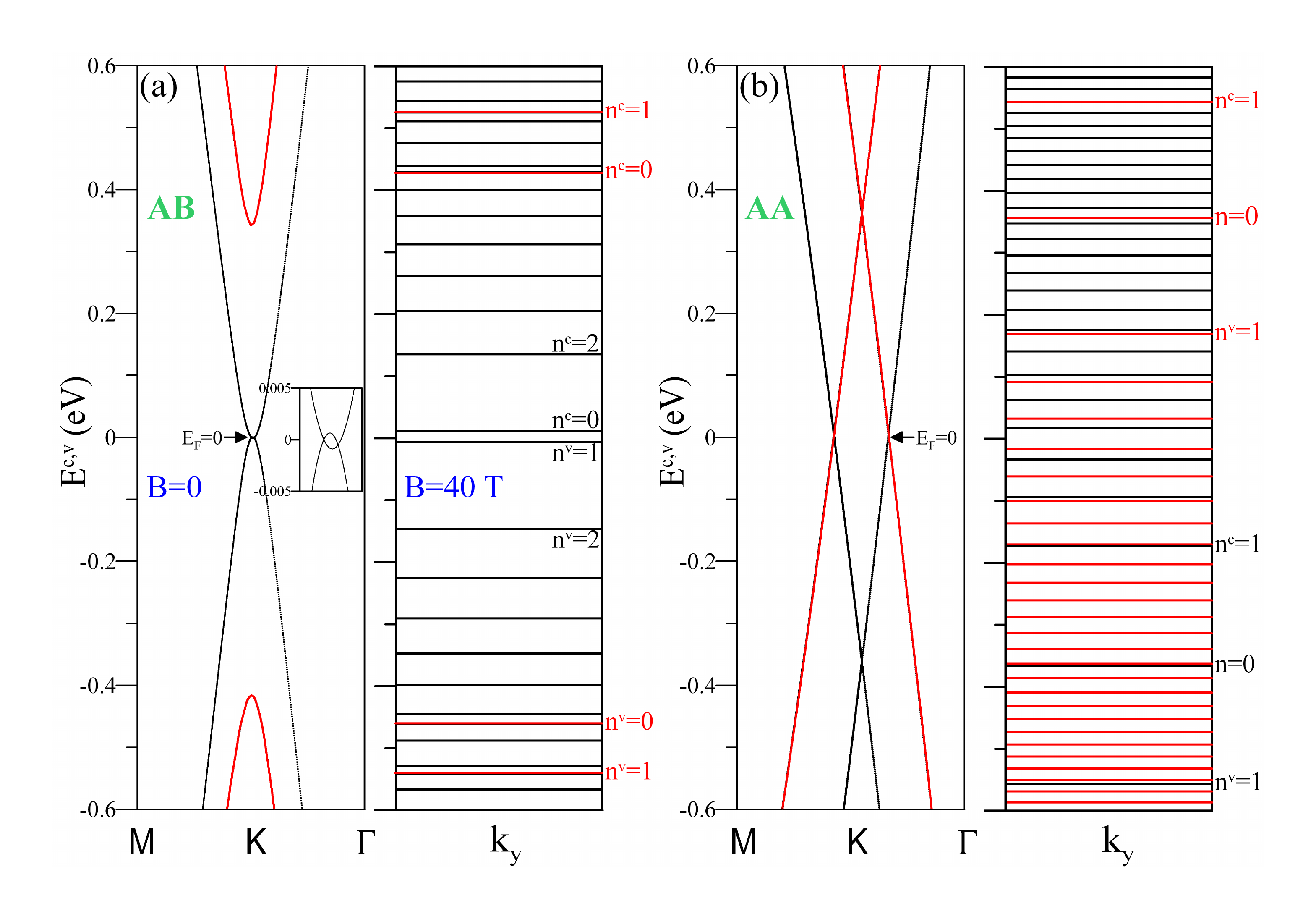}
  \end{center}
\end{figure}

\newpage

\begin{figure}
  \includegraphics[height=\textheight]{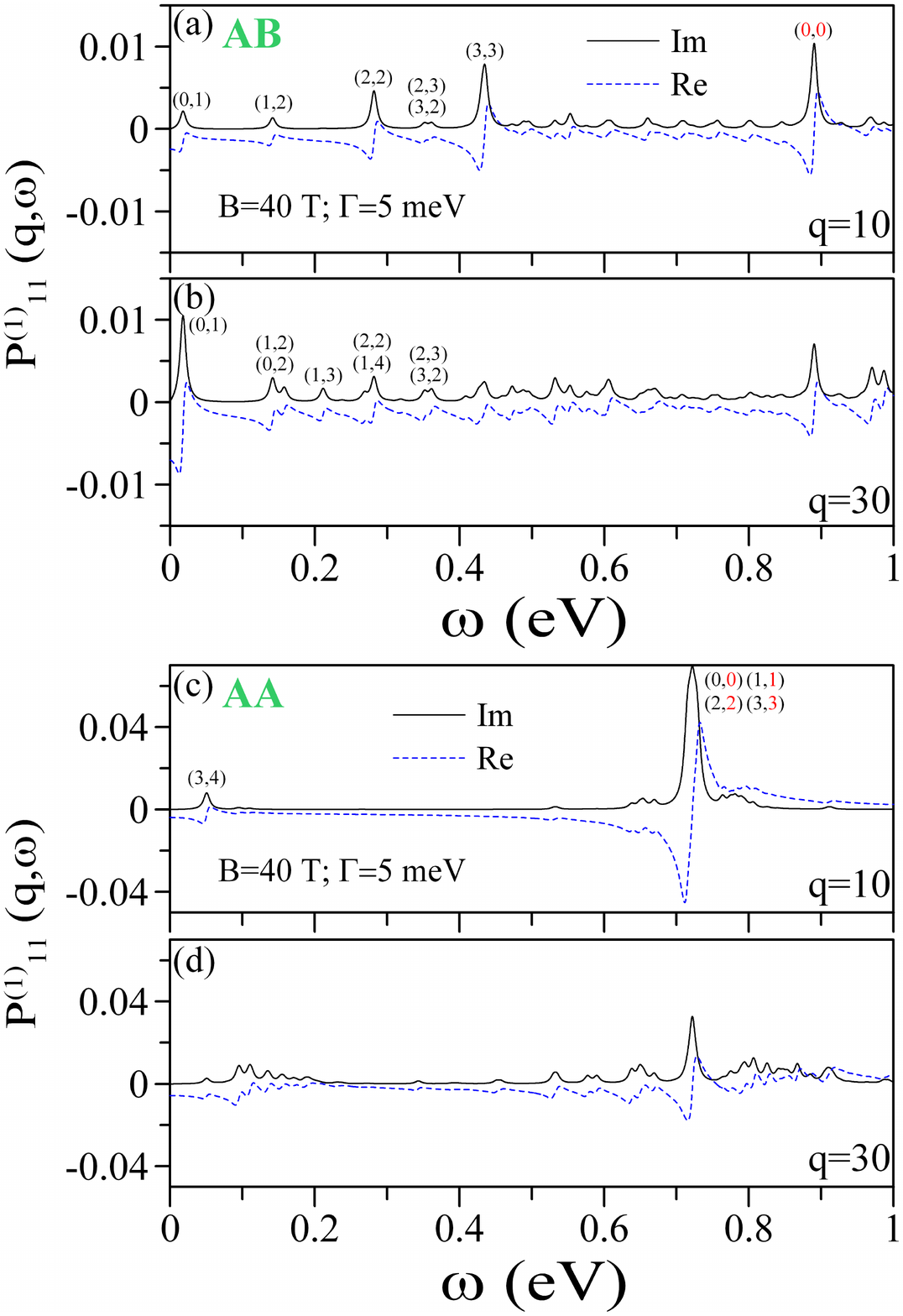}
\end{figure}

\newpage

\begin{figure}
  \includegraphics[height=\textheight]{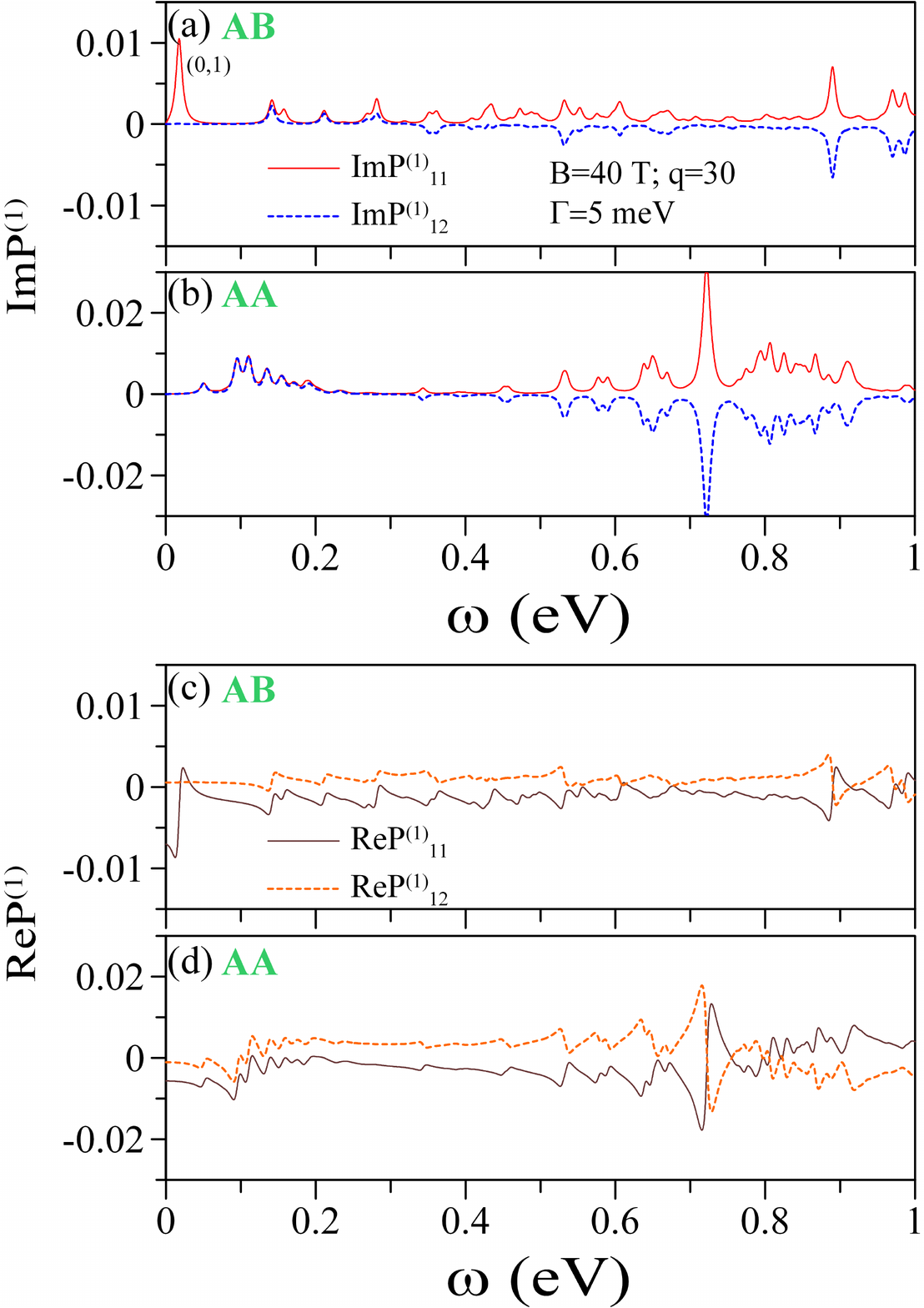}
\end{figure}

\newpage

\begin{figure}
  \includegraphics[height=\textheight]{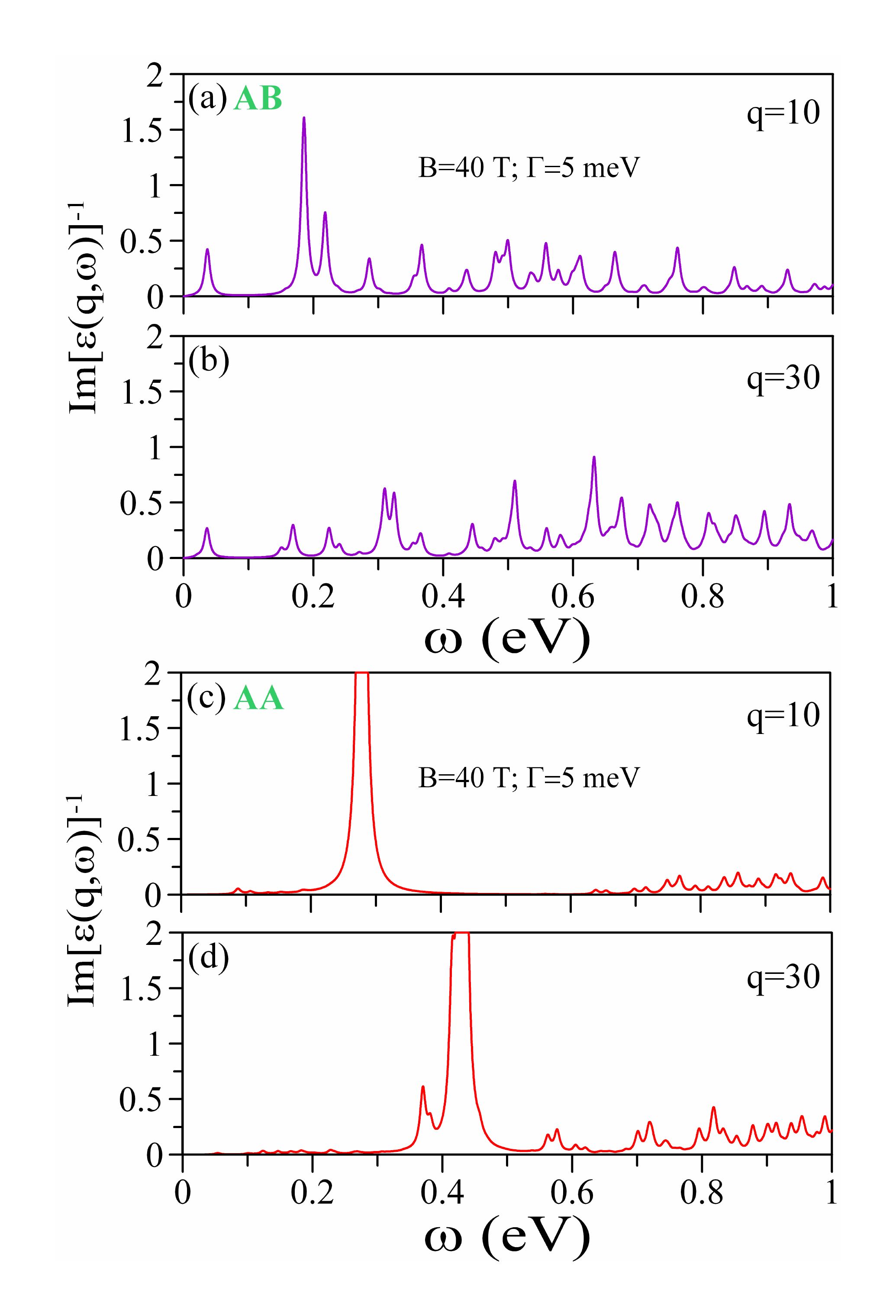}
\end{figure}

\newpage

\begin{figure}
  \includegraphics[height=\textheight]{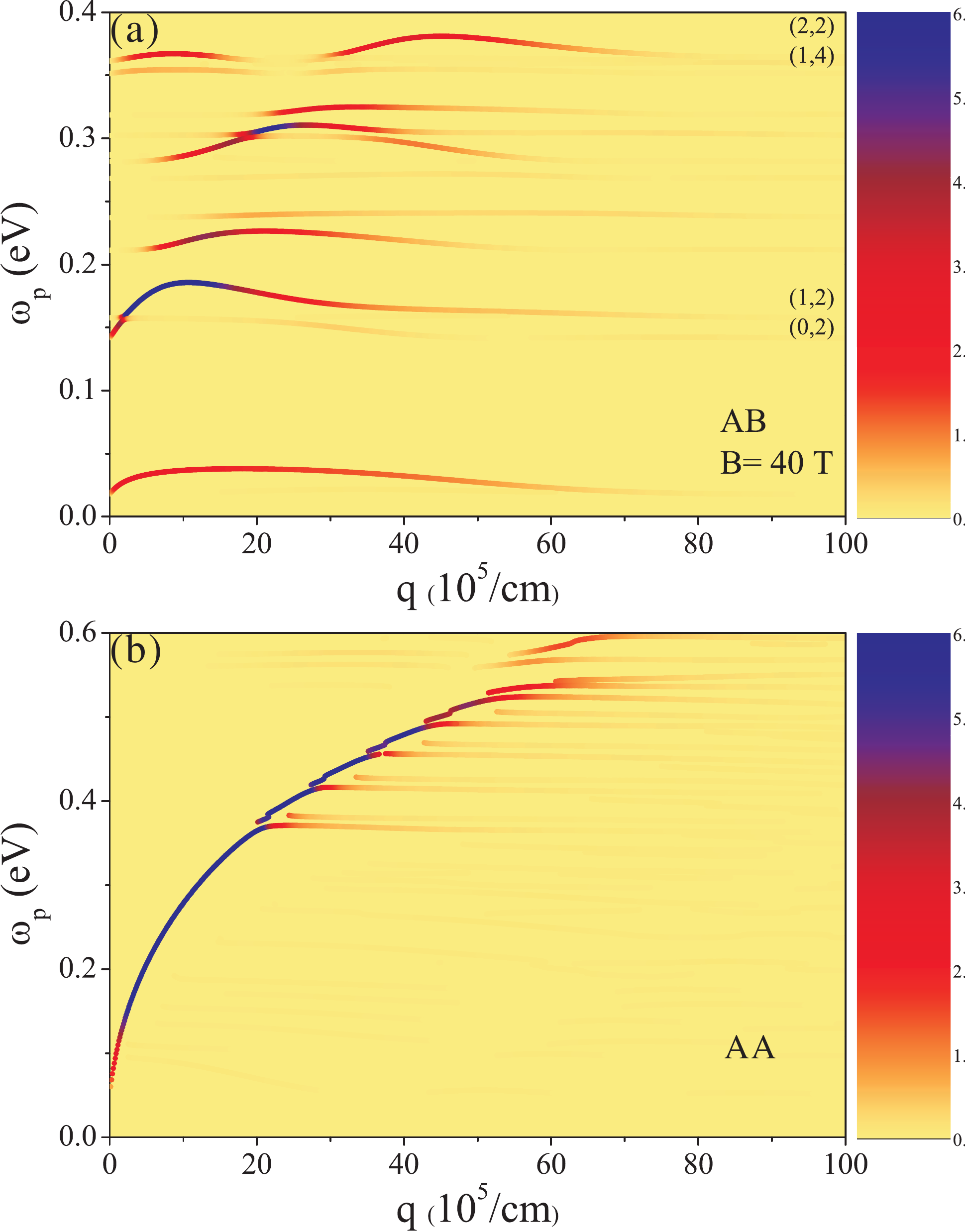}
\end{figure}

\newpage

\begin{figure}
  \includegraphics[height=\textheight]{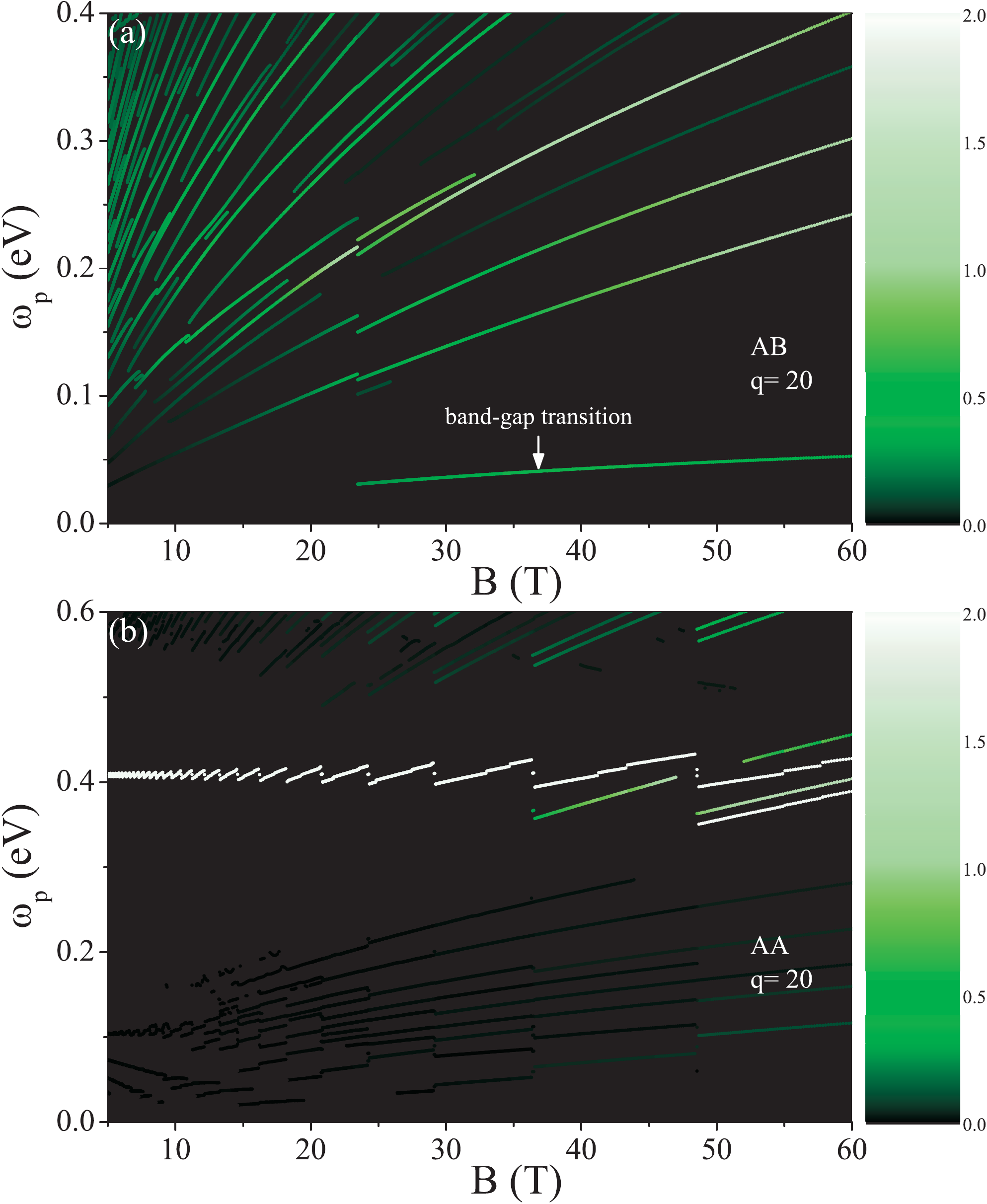}
\end{figure}

\end{document}